\documentclass{PoS}

\title{Restoration of Chiral Symmetry from a Boundary}

\ShortTitle{Restoration of Chiral Symmetry from a Boundary}

\author{
\speaker{B.~C.~Tiburzi}
\thanks{Work supported in part by a joint CCNY--RBRC fellowship, 
an award from the Professional Staff Congress of the CUNY, 
the Alfred P.~Sloan foundation through
a CUNY Junior Faculty Research Award in Science and Engineering,
and by the U.S.~National Science Foundation, 
under Grant No.~PHY-$1205778$. 
}\\
        Department of Physics,
        The City College of New York,  
        New York, NY, USA\\
        Graduate School and University Center,
        The City University of New York,
        New York, NY, USA\\
        RIKEN BNL Research Center, 
        Brookhaven National Laboratory, 
        Upton, NY, USA\\
        E-mail: \email{btiburzi@ccny.cuny.edu}}

\abstract{The imposition of Dirichlet boundary conditions in lattice computations obstructs the formation of a chiral condensate. We use chiral perturbation theory and meson models to address the effect of a Dirichlet boundary on chiral symmetry breaking. While pions are the longest-range modes in QCD, the restoration of chiral symmetry due to a boundary is shown not to depend upon the pion Compton wavelength but rather on that of the sigma meson. Power-law finite size corrections are exposed, and require prohibitively large lattices to overcome. We further speculate on the frustration of the chiral condensate for the case of confinement to the surface of a sphere.}

\FullConference{31st International Symposium on Lattice Field Theory - LATTICE 2013\\
		July 29 - August 3, 2013\\
		Mainz, Germany}

\begin{document}

%%%%%%%%%%
\section{Overview}

This talk concerns consequences of imposing fixed boundary conditions on the quark fields in QCD. 
Such boundary conditions present an obstruction in the formation of the chiral condensate; 
and, 
as a result, 
can lead to restoration of chiral symmetry. 
The issue of symmetry restoration from a boundary is an amusing topic in its own right, 
however, 
there is also a practical side to the matter. 
An increasing number of lattice QCD simulations employ fixed boundary conditions on the quark fields for various computational reasons. 
Consequently
it is useful to have in mind the possibility of restoring chiral symmetry, 
and to have a tool with which one can address this finite-size effect quantitatively. 
This is the main result presented here, 
based on the work in~\cite{Tiburzi:2013vza}.
We also provide a few speculations about modeling confinement in light of our study of Dirichlet boundary conditions and chiral symmetry breaking.

%%%%%%%%%%%%%%%%%%%%%%%%%%
\section{Chiral Perturbation Theory and a Boundary}

Spontaneous breaking of chiral symmetry is central to the study of low-energy QCD. 
Throughout we consider QCD with two light quark flavors, 
and in the limit that their masses are small,
$m_u, m_d \ll \Lambda_{QCD}$. 
In the massless limit, 
the QCD action has a chiral symmetry of the form 
$SU(2)_L \otimes SU(2)_R$
which is broken down to the vector subgroup
by the formation of the chiral condensate, 
$\langle \overline \psi_{j R} \psi_{i L} \rangle = - \lambda \delta_{ij}$. 
This quantity provides an order parameter for chiral symmetry. 
Fluctuations of the vacuum can be described by promoting the vacuum configuration to a field, 
$\delta_{ij} \to U_{ij}$, 
where the field 
$U$
lives in the coset 
$SU(2)_L \otimes SU(2)_R / SU(2)_V$. 
A simple parameterization is 
$U = e^{ i \vec{\pi} \cdot \vec{\tau} / v } = 1 + \frac{i \vec{\pi} \cdot \vec{\tau}}{v} + \cdots$, 
where 
$\vec{\pi}$
describes the iso-triplet of Goldstone pions.

Provided the parameter 
$v$
(which turns out to be the chiral-limit value of the pion decay constant)
is large, 
the fluctuations are Gau\ss ian, 
and can be described by a phenomenological low-energy effective theory, 
which is chiral perturbation theory. 
Within this theory, 
the effect of an explicit source of chiral symmetry breaking, 
such as a degenerate quark mass 
$m_q$, 
is described at lowest order by the term
\begin{equation}
\mathcal{L}
= 
- \frac{1}{4} m_q \lambda 
\, Tr
\left( U + U^\dagger
\right)
=
- \lambda m_q
\left( 
1 
- 
\frac{1}{2 v^2} 
\vec{\pi} \cdot \vec{\pi} 
+ \cdots
\right)
\label{eq:LO}
.\end{equation}
Expanding this term to first order gives rise to the vacuum energy density 
$ - \lambda m_q$, 
from which we can identify the chiral limit value of the chiral condensate, 
$\langle \overline \psi \psi \rangle = - \lambda$.
At second order, 
we have a mass term for the pions, 
$m_\pi^2 = \lambda m_q / v^2$.

Chiral perturbation theory can be used to address long-range physics in QCD. 
As such, 
the restoration of chiral symmetry on a compact manifold without a boundary has been treated some time ago in a series of papers~\cite{Gasser:1986vb}. 
One example is the finite temperature chiral transition, 
which arises due to the compactification of the Euclidean time direction. 
Another example is the restoration of chiral symmetry at finite volume, 
with the pion fields subject to periodic boundary conditions in space. 
At vanishing quark mass, 
chiral symmetry is not spontaneously broken in finite volume. 
One can glimpse the culprit of the effect by looking at the one-loop correction to the chiral condensate
\begin{equation}
\int \frac{d^4 k}{(2 \pi)^4}
\frac{1}{k^2 + m_\pi^2}
\sim 
m_\pi^2 \log m_\pi^2
\longrightarrow
\frac{1}{L^2} 
\left[ 
\frac{1}{( m_\pi L)^2} 
+ 
\sum_{n_\mu \neq 0}^\infty 
\frac{1}{ 4  \pi^2  n_\mu n_\mu + ( m_\pi L)^2} 
\right]
.\end{equation}
The infinite volume correction is a chiral logarithm, 
which vanishes in the chiral limit. 
On the other hand, 
the finite volume expression suffers a singularity in the chiral limit. 
This singularity points to the necessity of treating zero-momentum modes non-perturbatively, 
and the net effect is to restore chiral symmetry by averaging the zero mode over all directions in the coset manifold.

Here we consider the effect of fixed boundary conditions on the condensate. 
These boundary conditions frustrate the formation of the chiral condensate, 
$\langle \overline \psi \psi \rangle \big|_{boundary} = 0$. 
Boundary conditions leading to this behavior have been and continue to be imposed in lattice QCD calculations.

\begin{itemize}

\item \emph{Chopping of lattices}: temporal Dirichlet boundary conditions have been used as a time-saver to invert quark propagators on lattices of half their actual extent~\cite{Edwards:2005ym}.

\item \emph{Discontinuous external fields}: na\"ive inclusion of a uniform external field leads to 
boundary gradients, e.g.~the vector potential
$\vec{A} = - B y \, \hat{x}$
produces a magnetic field of the form
$\vec{B} = B \, \hat{z} [ 1 - L \delta ( y - L)]$.
Dirichlet boundary conditions have been sought to mitigate effects from the boundary non-uniformity of such external fields~\cite{Fiebig:1988en}.

\item \emph{Rotating lattices}: consideration of QCD in rotating frames leads one to impose spatial Dirichlet boundary conditions to avoid edge effects~\cite{Yamamoto:2013zwa}.

\item \emph{Schr\"odinger functional representation}: imposes inhomogeneous Dirichlet boundary conditions in time that lead to the vanishing of the scalar quark bilinear at the boundary~\cite{Sint:1993un}. 
Chiral symmetry restoration is irrelevant for computing the renormalization of operators in the massless Schr\"odinger functional scheme, 
however, 
hadron properties, such as those calculated in~\cite{Guagnelli:1999zf},
are subject to effects from the frustration of the chiral condensate.

\end{itemize}

\noindent
With these various applications, 
it makes sense to address quantitatively the effect of a Dirichlet boundary on the chiral condensate. 
To this end, 
we choose to impose a Dirichlet boundary in one direction, 
which we label "$x$",  
namely
$\psi (x=0) = \psi (x=L) = 0$.
Other directions are treated implicitly in this notation, 
and it will be irrelevant whether $x$ is a spatial or temporal direction.

As a consequence of the boundary conditions, 
the pion fields
$\vec{\pi} \sim \overline \psi \gamma_5 \vec{\tau} \psi$
vanish at the boundary, 
and the coset field 
$U$
is unity at the boundary. 
The latter actually poses a major problem in the computation of the chiral condensate. 
In chiral perturbation theory, 
the condensate is determined from 
Eq.~(\ref{eq:LO})
by the expression
\begin{equation}
\langle \overline{\psi} \psi (x) \rangle 
= 
- \frac{\lambda}{4} \langle U(x) + U^\dagger(x) \rangle
+ 
\cdots
\label{eq:cond}
,\end{equation}
where further terms are functions of 
$U(x)$ 
and its derivatives. 
Due to the boundary behavior of the coset field, 
the condensate calculated by 
Eq.~(\ref{eq:cond}) 
will never vanish at the boundary. 
This is merely an indication that the coset field cannot fluctuate off the manifold. 
To address chiral symmetry restoration in the presence of the Dirichlet boundary, 
we require degrees of freedom beyond pions. 
Adding further degrees of freedom changes the description from universal to model dependent.  
As such, 
the imposition of a Dirichlet boundary potentially gives one a probe of the dynamics underlying spontaneous chiral symmetry breaking.

%%%%%%%%%%%%
\section{Sigma Model}

Here we use a very simplified model of spontaneous chiral symmetry breaking by including the sigma meson degree of freedom. 
In the sigma model, 
the sigma field condenses to break the chiral symmetry spontaneously. 
The Euclidean action density for the model is given by
\begin{equation}
\mathcal{L}
= 
\frac{1}{2} 
\partial_\mu S \partial_\mu S
+ 
\frac{1}{2} 
\partial_\mu \vec{P} \cdot \partial_\mu \vec{P}
- 
\frac{\lambda m_q}{v} S
+
\Lambda ( S^2 + \vec{P} \, {}^2 - v^2)^2 
.\end{equation}
The model has an 
$SO(4)$
global symmetry that is spontaneously broken to 
$SO(3)$
by the vacuum expectation value of 
$S$
and
$\vec{P}$
fields, 
namely 
$S_0^2 + \vec{P} \, {}^2_0 = v^2$. 
This model has a long history and can be derived from the NJL model, see~\cite{Ebert:1985kz}.
The connection between the Casimir effect and spontaneous symmetry breaking in the NJL model has been recently investigated~\cite{Flachi:2012pf}.

To investigate the sigma model, 
it is convenient to introduce a polar decomposition of the fields, 
$S + i \vec{P} \cdot \vec{\tau} \equiv \Sigma U$, 
where 
$\Sigma$ 
is a real field, 
and 
$U$ 
an 
$SU(2)$-valued field. 
In terms of polar variables, 
the sigma model action takes the form
\begin{equation}
\mathcal{L} 
=
\frac{1}{4} Tr \, \left[ \partial_\mu \Sigma \partial_\mu \Sigma + \Sigma^2 \partial_\mu U \partial_\mu U^\dagger \right]
- 
\frac{\lambda m_q}{4 v} Tr \, \left[ \Sigma ( U + U^\dagger) \right]
+ 
\Lambda ( \Sigma^2 - v^2)^2
,\end{equation}
and has an 
$SU(2) \otimes SU(2)$
symmetry in the chiral limit that is spontaneously broken to 
$SU(2)$
when the fields pick up their vacuum expectation values, 
$U_0 = 1$ 
and
$\Sigma_0 = v$. 
Expanding these fields about their vevs, 
we find the Gell--Mann-Oakes-Renner relation, 
$m_\pi^2 = \lambda m_q / v^2$, 
and a relation for the mass of the sigma meson
$m_\sigma^2 = 8 \Lambda v^2$. 
This enables us to fix the parameter 
$\Lambda$
from phenomenology~\cite{Caprini:2005zr}.

Imposing Dirichlet boundary conditions on the quark fields turns into boundary conditions on the meson fields of the sigma model. 
As before, 
the unitary field satisfies the unit boundary conditions
$U( x = 0) = U (x = L) = 1$, 
while the sigma field satisfies Dirichelt boundary conditions, 
$\Sigma(x = 0) = \Sigma(x=L) = 0$. 
Computation of the chiral condensate from the sigma model,
\begin{equation}
\langle \overline{\psi} \psi (x) \rangle 
= 
- \frac{\lambda}{4 v} \Sigma(x) 
\left\langle
U(x) + U^\dagger(x) 
\right\rangle
\end{equation}
will consequently satisfy the correct boundary conditions, 
compare with Eq.~(\ref{eq:cond}). 
The vacuum expectation value of the sigma field is now generally coordinate dependent, 
$\Sigma_0 = \Sigma_0(x)$, 
and can be determined by minimizing the Euclidean action
\begin{equation}
S[\Sigma_0]
=
\int_0^L dx \, 
\left[ 
\frac{1}{2} \left( \frac{d \Sigma_0}{dx} \right)^2 + \Lambda ( \Sigma_0^2 - v^2)^2 
\right]
,\end{equation}
subject to the fixed endpoints, 
$\Sigma_0(0) = \Sigma_0(L) = 0$. 
Finding the condensate is an exercise in analytical mechanics that turns out to be soluble in terms of elliptic integrals.

%%%%%%%%%%%%%%%%%%%%%%%%%%%%%%%%%%%%%%
%
%%%%%%%%%%%%%%%%%%%%%%%%%%%%%%%%%%%%%%
%
%%%%%%%%%%%%%%%%%%%%%%%%%%%%%%%%%%%%%%
\begin{figure}
\begin{center}
\includegraphics[width=0.45\textwidth]{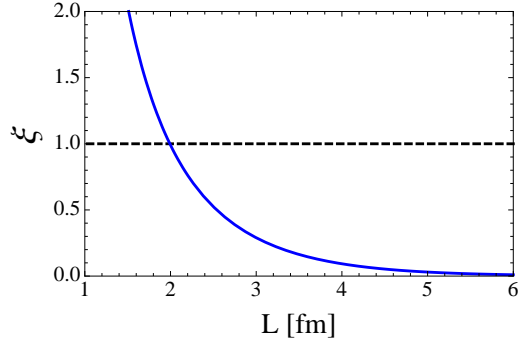}
\caption{
The solution for the mechanical analogue of energy
$\xi$ 
as a function of the finite extent 
$L$ 
of the 
$x$-direction. 
Values of 
$L$ 
for which
$\xi \geq 1$
lead to the vacuum expectation value
$\Sigma_0(x) = 0$, 
and hence correspond to a complete restoration of chiral symmetry in the sigma model.
}
\label{f:xi}
\end{center}
\end{figure}
%%%%%%%%%%%%%%%%%%%%%%%%%%%%%%%%%%%%%
%
%%%%%%%%%%%%%%%%%%%%%%%%%%%%%%%%%%%%%
%
%%%%%%%%%%%%%%%%%%%%%%%%%%%%%%%%%%%%%

The solution of the equation of motion for 
$\Sigma_0$
has at most one turning point. 
While there are solutions with multiple turning points, 
the value of 
$\Sigma_0$
will necessarily change sign, 
and the inclusion of a small quark mass will raise the action of such solutions. 
Hence the action is minimized for a solution with one turning point located at
\begin{equation}
max \, \, \Sigma_0 = v \sqrt{1 - \xi}
,\end{equation}
where 
$\xi$
is the analogue of energy for the system. 
The energy can be determined as a function of the length
$L$, 
see Figure \ref{f:xi},
and must satisfy 
$\xi < 1$. 
When 
$\xi \geq 1$, 
the value of the sigma condensate is zero everywhere.

\begin{figure}
\begin{center}
\includegraphics[width=0.45\textwidth]{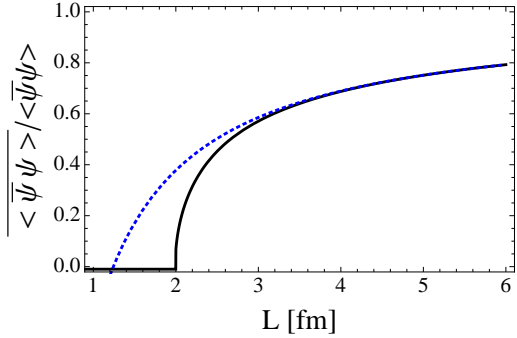}
\caption{
Ratio of the volume-averaged condensate 
$\overline{\langle \overline{\psi} \psi \rangle }$ 
to the infinite volume condensate
$\langle \overline{\psi} \psi \rangle$
plotted as a function of the finite extent 
$L$. 
The dotted curve shows the asymptotic formula
Eq.~(3.7).
}
\label{f:power}
\end{center}
\end{figure}

From direct integration of the equations of motion, 
one can implicitly determine 
$\Sigma_0$ 
as a function of 
$x$.
In turn, 
this information can be used to determine the chiral condensate
$\langle \overline{\psi} \psi (x) \rangle$, 
which properly vanishes at the boundary. 
In the bulk of the lattice, 
a non-zero value of the condensate can form provided 
$L \gtrsim 2 \, \texttt{fm}$. 
The maximum value of the condensate is located at the turning point, 
namely at 
$x = \frac{L}{2}$. 
At this point, 
the value of the condensate swiftly approaches the infinite volume value. 
Analytically we find the asymptotic formula
$\langle \overline{\psi} \psi (\frac{L}{2}) \rangle / \langle \overline{\psi} \psi \rangle = 
1 - 4 \exp ( - \frac{1}{2} m_\sigma L)$. 
If one can locate the physics near the midpoint 
(such as in simulations utilizing temporal Dirichlet boundary conditions), 
the effect on the condensate is not too dramatic. 
For spatial Dirichlet boundary conditions,  
however, 
a more natural measure of the finite size effect is the volume-averaged condensate
\begin{equation}
\overline{ \langle \overline{\psi} \psi \rangle}
\equiv
\frac{1}{L} \int_0^L dx \, \langle \overline{\psi} \psi (x) \rangle
.\end{equation}
The behavior of the volume-averaged condensate is shown in Figure \ref{f:power} as a function of 
$L$. 
Considerable finite-size effects are seen, 
and analytically we can show the approach to infinite volume is only power law
\begin{equation}
\overline{ \langle \overline{\psi} \psi \rangle}
/
\langle \overline{\psi} \psi \rangle
=
1
- 
\frac{4 \log 2}{ m_\sigma L}
\label{eq:asym}
.\end{equation}
Notice the dependence on the Compton wavelength of the sigma meson instead of the pion. 
In this case, 
the long-range physics is controlled by a mode which is not the most infrared. 
Finally we must stress that the behavior of the condensate has been determined in a model-dependent fashion. 
There are various avenues one can take to improve the model. 
Higher-lying scalar states have a rather unusual spectroscopy. 
These should be included in a more detailed study, 
as they play a role in the mechanism underlying chiral symmetry breakdown, 
see~\cite{Fariborz:2008bd}, and references therein.

%%%%%%%%%%%%%%%%%%%%%%%%%%%%%%%%%%%%%%
%
%%%%%%%%%%%%%%%%%%%%%%%%%%%%%%%%%%%%%%
%
%%%%%%%%%%%%%%%%%%%%%%%%%%%%%%%%%%%%%%
%%%%%%%%%%%%%%%%%%%%%%%%%%%%%%%%%%%%%
%
%%%%%%%%%%%%%%%%%%%%%%%%%%%%%%%%%%%%%
%
%%%%%%%%%%%%%%%%%%%%%%%%%%%%%%%%%%%%%

%%%%%%%%%%%%%%%%%
\section{Bag Model Confinement}

In the remaining (space)time, 
we shall present some speculations concerning the relation between bag-model confinement 
and chiral symmetry breaking in light of our study of the sigma model. 
The MIT bag model~\cite{Chodos:1974je} enforces a boundary condition on the quark fields, 
having the form
$\vec{n} \cdot \overline{\psi}  \, \vec{\gamma} \psi \big|_{R(\theta,\phi)} = 0$,
which ensures the confinement of color current in the model. 
We shall take the spatial surface
$R(\theta, \phi)$
to be a sphere.
One can show that such boundary conditions imply the vanishing of the scalar bilinear on the surface of the bag, 
$\overline{\psi} \psi \big|_R = 0$; 
and, 
therefore, 
the vanishing of the chiral condensate
$\langle \overline{\psi} \psi \rangle \big|_R = 0$.

Within the sigma model,
we can enforce the bag boundary condition by having the scalar meson field vanish on the surface of a sphere of radius 
$R$. 
One question that can be asked is whether a non-zero chiral condensate can develop inside the bag. 
To address this question, 
one minimizes the action for the vev of the sigma field
\begin{equation}
S[\Sigma_0]
=
\int d \vec{r} \, 
\left[
\frac{1}{2} \vec{\nabla} \Sigma_0 \cdot \vec{\nabla} \Sigma_0 
+ 
\Lambda ( \Sigma_0^2 - v^2)^2
\right] 
.\end{equation}
Angular gradients contribute positively to the action and can be eliminated by seeking spherically symmetric solutions. 
After rescaling the field, 
$x \equiv \frac{1}{v} m_\sigma r \Sigma_0$
and renaming
$t \equiv m_\sigma r$, 
the problem maps into the classical dynamics of a time-dependent force
\begin{equation}
\ddot{x}
= 
\frac{x}{2} \left( \frac{x^2}{t^2} - 1 \right)
.\end{equation}
Numerically there only appears to be a trivial solution to this equation, 
$x(t) = 0$, 
corresponding to chiral symmetry restoration. 
What is surprising is that this remains true no matter how large 
$R$ 
is taken to be. 
So far we have not been able to find a simple analytic argument as to why. 
The sigma model points to an inability to confine chiral symmetry breaking to within a sphere.

Another question that can be posed concerns what happens outside the bag. 
Does a chiral condensate exist far from the bag surface?
Within the sigma model, 
the large radius behavior of the sigma vev must have the form
%\begin{equation}
$\Sigma_0 (r) \sim \frac{v}{m_\sigma r} \exp \left( \pm \frac{ i m_\sigma r}{ \sqrt{2}} \right)$,
%\end{equation} 
and hence there is no solution with a uniform condensate at infinity. 
Consequently it is not possible to break chiral symmetry outside the bag. 
Furthermore as chiral symmetry is restored inside the bag, 
another way to phrase our observation is that
the sigma model does not support the existence of spherical droplets of chirally symmetric matter. 
Perhaps this is not surprising if we remind ourselves that the chiral phase transition is second order in the sigma model. 
If the model were to support chirally symmetric spherical droplets as solutions, 
we might expect a first-order phase transition.

%%%%%%%%%%
\section{Summary}

The use of Dirichlet boundary conditions can obviously alter the phase of QCD. 
Unlike periodic boundary conditions, 
the resulting finite volume effects are power law in nature;
and, 
surprisingly, 
are not controlled by the pion Compton wavelength. 
We argue that the sigma meson plays a crucial role in the restoration of chiral symmetry from a boundary, 
even though the pions are the most infrared modes of the theory. 
For lattice QCD calculations using spatial Dirichlet boundary conditions, 
there is good cause to be concerned about such effects unless lattices are considerably longer than several 
$\texttt{fm}$. 
The burden of proof ultimately relies with practitioners of these lattice methods. 
For calculations using temporal Dirichlet boundary conditions, 
one can localize the physics in the bulk of the lattice, 
but one still should exercise caution concerning finite temporal effects. 
In the course of our study, 
we find that bag confinement and chiral symmetry breaking appear to be incompatible. 
Here there is a curious dependence on the number of dimensions 
and on the boundary geometry. 
We suggest that the inability to form spherical droplets that are chirally symmetric 
is linked with the order of the chiral phase transition in the sigma model.   
It would be quite interesting to study this direction further.

\end{document}